\documentclass[pra,aps,showpacs,superscriptaddress,twocolumn]{revtex4}
\usepackage{amsmath}
\usepackage{amsfonts}
\usepackage{amssymb}
\usepackage{mathrsfs}
\usepackage{epsfig}
\usepackage{bm}
\usepackage{amsbsy}
\usepackage{color}
\usepackage{epstopdf}
\usepackage{txfonts}

\newcommand{\ej}[1]{E_\mathrm{J_{#1}}}
\newcommand{\ec}[1]{E_\mathrm{C_{#1}}}
\newcommand{\eel}[1]{E_\mathrm{el_{#1}}}

\def\be{\begin{equation}}
\def\ee{\end{equation}}

\begin{document}


\title{Non-Markovian qubit dynamics in a circuit-QED setup}

\author{P. C. C\'ardenas}
\affiliation{Centro de Ci\^encias Naturais e Humanas, Universidade Federal do ABC, 09210-170, Santo Andr\'e, S\~ao Paulo, Brazil}
\author{M. Paternostro}
\affiliation{Centre for Theoretical Atomic, Molecular and Optical Physics,
School of Mathematics and Physics, Queen's University, Belfast, BT7 1NN, United Kingdom}
\author{F. L. Semi\~ao}
\affiliation{Centro de Ci\^encias Naturais e Humanas, Universidade Federal do ABC, 09210-170, Santo Andr\'e, S\~ao Paulo, Brazil}
%
\begin{abstract}
We consider a circuit-QED setup that allows the induction and control of non-Markovian dynamics of a qubit. Non-Markovianity is enforced over the qubit by means of its direct coupling to a bosonic mode which is controllably coupled to other qubit-mode system. We show that this configuration can be achieved in a circuit-QED setup consisting of  two initially independent superconducting circuits, each formed by one charge qubit and one transmission-line resonator, which are put in interaction by coupling the resonators to a current-biased Josephson junction. We solve this problem exactly and then proceed with a thorough investigation of the emergent non-Markovianity in the dynamics of the qubits. Our study might serve the context for a first experimental assessment of non-Markovianity in a multi-element solid-state device. 
\end{abstract}

\maketitle

The modeling of real physical systems in terms of  bosonic and two-level systems, and their interaction, is ubiquitous in physics, especially in optical or atomic scenarios~\cite{Haroche}. A well-known example is the celebrated Jaynes-Cummings model describing the interaction of a two-level atom with a single mode of the quantized electromagnetic field in the rotating wave approximation  \cite{jcm}. Many interesting physical phenomena arise in this kind of system depending on details of the interaction between its parts. One may mention the appearance of quantum phase transitions \cite{qpt}, polariton physics \cite{martin_laserp}, and phononic non-linearities \cite{nl}, just to name a few. Interacting two-level atoms and bosonic modes provide also a natural (and somehow historical) route to address questions of quantum open-system dynamics~\cite{RMPCaldeira,book_dec} and may be applied to a great variety of systems ranging from superconductivity \cite{caldeira_legget} to chromophores in biological complexes \cite{bio_open}.

Modern circuit-QED setups are good examples of well controlled systems where the interaction between two-level systems and bosonic modes can be experimentally investigated \cite{N,Schoelkopf,cqed,3}. They involve manipulation and control of the interaction between superconducting circuits behaving as artificial atoms (two-level systems) and one-dimensional transmission-line resonators (bosonic systems). The low dissipation and the small mode volume of the circuits together with big effective dipole moments of the superconducting qubits favors the achievement of the strong-coupling regime, where quantum behavior can be observed \cite{N,3,cqed}. 

On the other hand, memory effects in open systems is another topic of great interest \cite{general,quant}. Memoryless or Markovian evolutions represent a limited portion of the rich scenario of open system dynamics.  Much effort has then been directed to characterize, quantify and manipulate the degree of non-Markovianity of physical systems \cite{quant}. In this work, we propose a circuit-QED scheme to study the emergence of non-Markovianity in the dynamics of the qubits, tuning the details of the evolution by exploiting the great flexibility of the setup that we address. We start in Sec. I by presenting the system and solving the associated model exactly. We then move to Sec. II, where we evaluate the degree of non-Markovianity for the qubits in the case of modes initially prepared in coherent states. Sec. III addresses the role that phase 
coherence has in the phenomenology highlighted here by studying the case of phase diffused coherent states. We draw our conclusions in Section IV.
%
%
\section{Model and Setup}\label{model}
\begin{figure}[htb]
\centering\includegraphics[width=1\columnwidth]{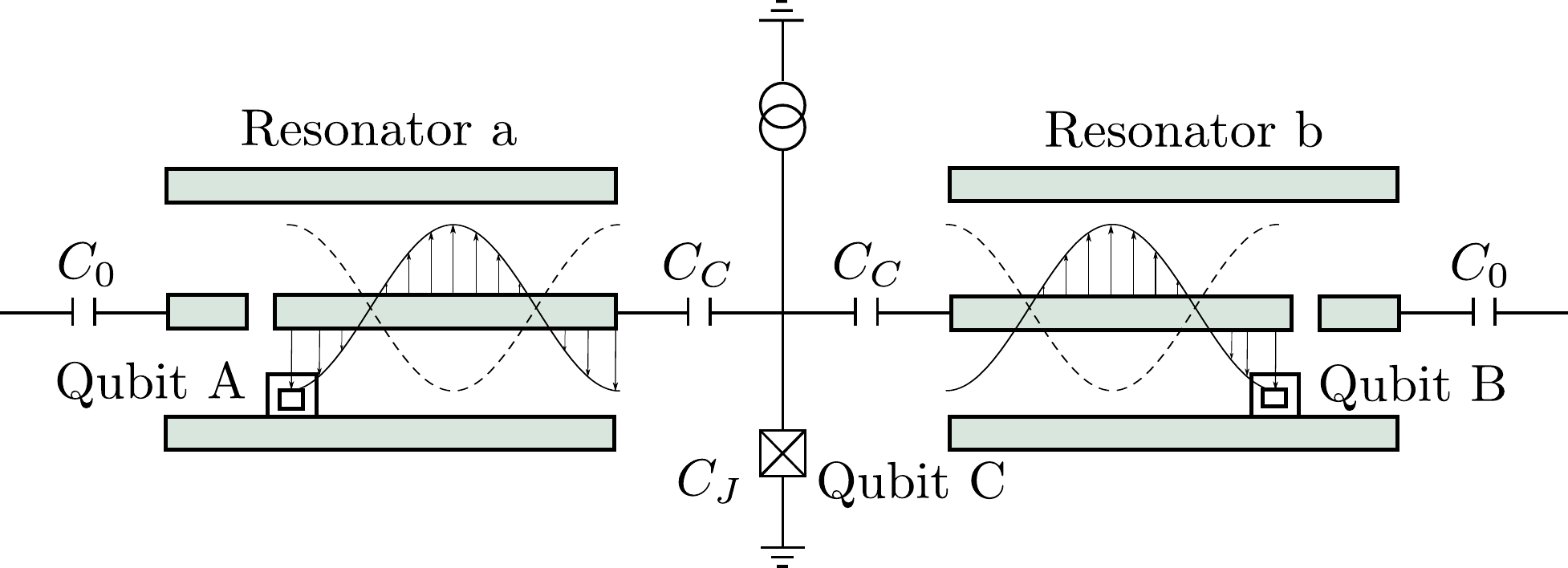}
\vspace{-0.5cm}
\caption{(Color
  online) Sketch of the elements comprising the circuit-QED model used in this work. Charge qubit $A$ is coupled to a  transmission line resonator $a$ while charge qubit $B$ is coupled another transmission line resonator $b$. These are Cooper pair boxes with tunable Josephson coupling. The coupling between the resonators is achieved by dispersively coupling a CBJJ qubit $C$ to both resonators.}   
\label{setup}       
\end{figure}

The circuit-QED system we propose for studying the effect of cross couplings of localized modes in spin-boson systems is depicted in Fig.\ref{setup}. It combines two different circuit-QED setups in a single versatile setup. On one hand, Cooper pair box qubits with tunable Josephson coupling \cite{caldeira_legget} are capacitively coupled to different single mode high-Q superconducting coplanar resonators $a$ and $b$ (frequencies $\omega_a$ and $\omega_b$). This forms two local and non interacting spin-boson systems. Each of these local circuits are essentially the well developed setups used in many experiments involving superconducting qubits and transmission line resonators \cite{Schoelkopf}. It is important to emphasize that today there are improved noise robust superconducting qubit architectures, like the transmon \cite{transmon}, but the basic elements of our proposal do not depend strongly on the specific type of qubit. Of course, in practice, the more protected the design is against decoherence the better for observing quantum coherence effects.

We employ a third qubit $C$, now a current biased Josephson junction (CBJJ), in order to dispersively induce an indirect coupling between the resonators. These kind of qubit is specially suited for this task since small changes in the bias current can strongly change the detuning with the resonators. This has been shown in \cite{guo} and employed in \cite{njp} to implement single qubit operations in linear optics quantum computing using circuit-QED. Coupled modes interacting with qubits are also the basic elements in the Jaynes-Cummings-Hubbard lattices \cite{JCH}. However, it is important to notice that the form of the interaction term involving the fields and qubits in our case is not of the Jaynes-Cummings one, as discussed below. A similar setup have been proposed to generate entanglement between superconducting qubits using dynamical Casimir effect \cite{casimir}.  Coupled circuits have also been proposed to simulate particular quantum environments for superconducting qubits with the purpose of simulating exciton transfer in photosyntetic systems \cite{A}. In what follows, we pursuit a different line of investigation by considering how the coherent coupling of one of the superconducting qubits with a resonator mode, which is controlably coupled to another qubit-resonator system, compete with dephasing to bring about memory effects. These effects are then quantified by using modern quantum information tools and by exactly solving the dynamics as explained later on in this work.

The full Hamiltonian for this system is given by $H=H_0+H_{SB}+H_{BB}$ (we choose units such that $\hbar =1$ throughout the paper), with  
\begin{eqnarray}
H_0=\frac{\omega_A}{2}\sigma_{z,A}+\frac{\omega_B}{2}\sigma_{z,B}+\frac{\omega_C}{2}\sigma_{z,C}+\tilde{\omega}_aa^\dag a+\tilde{\omega}_bb^\dag b,
\end{eqnarray}
where $\omega_C$ is the transition frequency of the CBJJ qubit and $\omega_A(B) = \sqrt{\ej{A(B)}^2+\eel{A(B)}^2}$ the transition frequency of qubit $A(B)$. All these are determined by the electrostatic energy $\eel{A(B)} = 4\ec{A(B)}(1-2n_{g_{A(B)}})$ with $\ec{A(B)} = e^2/2C_{\Sigma_{A(B)}}$ being the charging energy, $\ej{A(B)} = E^\mathrm{max}_\mathrm{J_j}
\cos(\pi \Phi_j/\Phi_0)$ the Josephson coupling energy, $C_{\Sigma_{A(B)}}$ the total box capacitance, $n_{g_{A(B)}} = C_{g_{A(B)}}V_{g_{A(B)}}/2e$ the dimensionless gate charge, and $E^\mathrm{max}_\mathrm{J_{A(B)}}$ the maximum Josephson energy. Finally, $C_{g_{A(B)}}$ is the gate capacitance, $V_{g_{A(B)}}$ is the gate voltage, and $\Phi_{A(B)}$ is an externally applied flux (with $\Phi_0$ being the flux quantum). Any dispersive shift of the natural frequencies of the resonators due to coupling with qubit $C$ is absorbed into $\tilde{\omega}_i$ ($i=a,b)$~\cite{njp}. We now pass on to the discussion of the qubit-mode (spin-boson) Hamiltonian, which reads
\begin{equation}
\begin{aligned}
H_{SB}&=g_{a} \left( \mu_{A} - \cos\theta_{A} \sigma_{z,A} + \sin\theta_{A}\sigma_{x,A}  \right)(a^\dag+a)\\
&+g_{b} \left( \mu_{B} - \cos\theta_{B} \sigma_{z,B} + \sin\theta_{B}\sigma_{x,B}  \right)(b^\dag+b),
\end{aligned}
\end{equation}
where $\mu_{A(B)}=1-2n_{g_A(B)}$, $g_{a(b)} =
e(C_{g,{A(B)}}/C_{\Sigma,{A(B)}})V^0_{{\mathrm{rms},a(b)}}$ is the coupling strength of the interaction between qubit $A$ ($B$) and mode $a$ ($b$), $\theta_{A(B)} = \arctan[\ej{j}/\ec{j}(1-2n_{g,j})]$, and $V^0_\mathrm{rms,a(b)} = \sqrt{\omega_{a(b)}/2C_{a(b)}}$ is the rms value of the voltage in the ground state of resonator $a$ ($b$), with $C_{a(b)}$ being the total capacitance of  the transmission line $a(b)$. Finally, the boson-boson coupling Hamiltonian $H_{BB}$, which is indirectly induced by mutual coupling with qubit $C$, reads  \cite{guo}
\begin{eqnarray}
H_{BB}&=&\lambda(a^\dag b+b^\dag a),
\end{eqnarray}
where the value of the coupling strength $\lambda$ can be tuned by addressing the CBJJ (qubit $C$) and choosing a properly bias current \cite{guo}. From now on, we will drop $(\omega_C/2)\sigma_{z,C}$ from $H_0$ since we now assume this qubit to be prepared in an eigenstate of $\sigma_{z,C}$. 

Our goal is to engineer an effective interaction Hamiltonian for each local spin-boson system having the form $V_{Jj}\propto\sigma_{z,J}(j^\dag+j)$ for $j=a$ ($j=b$) if $J=A$ ($J=B)$. Consequently, we need $\theta_{j}=0$, a condition that can be achieved by imposing $n_{g,j}\neq1/2$ and by tuning the external flux on the charge qubits so as to satisfy the relation $\Phi_j=(k+1/2)\,\Phi_0$ with $k\in{\mathbb Z}$. Although in the remainder of the paper we will ensure to work \textit{very closely} to the charge degeneracy point $n_{g,j}=1/2$, we cannot be exactly at it, as this would lead to  $\cos\theta_{j}=0$. Working at the degeneracy point reduces the impact of dephasing on the qubit state~\cite{vion,dep}. Consequently, we will have to explicitly include dephasing for the qubits in the dynamical equation of motion which will be discussed later on.

By assuming identical resonators ($\tilde{\omega}_{a,b}=\omega$) as well as identical qubits ($n_{g_A(B)}=n_g$, $\omega_{A,B}=\omega_0$, and $g_A=g_B$), the full Hamiltonian becomes
\begin{equation}\label{Hf}
\begin{aligned}
H&=\frac{\omega_0}{2}(\sigma_{z,A}+\sigma_{z,B})+\omega (a^\dag a+ b^\dag b)+\lambda(a^\dag b+b^\dag a)\\
&+g(\sigma_{z,A}+\mu\openone_A)(a^\dag+a)+g(\sigma_{z,B}+\mu\openone_B)(b^\dag+ b).
\end{aligned}
\end{equation}
As a consequence of working slightly out of the degeneracy point ($\mu\neq 0$), we get driving-like terms on the modes proportional to $(a^\dag+a)$ and $(b^\dag+b)$.

We are now in a position to present the equations of motion for the system. 
As discussed before, we are not working exactly at the degeneracy point and then qubit dephasing should be taken into account \cite{dep}. Regarding other decoherence mechanisms, dissipation affecting the qubits or the transmission lines, as well as dephasing on the latter, can be made negligibly small compared to dephasing in the qubits~\cite{Schoelkopf}. Therefore, by keeping only the dominant terms, the dynamics will be governed by the master equation
\begin{eqnarray}\label{dep}
\frac{\partial\rho}{\partial t}=-i[H,\rho]+\frac{\gamma}{2}\sum_{J=A,B}\left(\sigma_{z,J}\rho\sigma_{z,J}-\rho\right)
\end{eqnarray}
with $\gamma$ the single-qubit dephasing rate and $H$ given by Eq.~(\ref{Hf}). 

Using the two-mode displacement operator $T=e^{\xi a^\dag-\xi^* a}\otimes e^{\xi b^\dag-\xi^* b}$ [with $\xi=\mu g/(\omega+\lambda)$] and the beam-splitter transformation $T'=\exp[\pi(a^\dag b-a b^\dag)/4]$~\cite{bst}, it is straightforward to derive the effective Hamiltonian model
\begin{equation}
\begin{aligned}
H''&=\frac{\omega'_0}{2}\left(\sigma_{z,A}+\sigma_{z,B}\right)+\omega_+a^\dag a+ \omega_-b^\dag b\\
&+\frac{g}{\sqrt{2}}[\sigma_{z,A}(a^\dag+a-b^\dag-b)+\sigma_{z,B}(b^\dag+b+a^\dag+a)],
\end{aligned}
\end{equation}
where $\omega'_0=\omega_0-4g\xi$, $\omega_\pm=\omega\pm\lambda$, and no direct interaction between the field modes is present in this picture. It is now possible to decouple 
the spin and boson degrees of freedom using the polaron transformations~\cite{pt}
\begin{equation}
T''=\exp[\lambda_+(\sigma_{z,B}+\sigma_{z,A})(a^\dag-a)+\lambda_-(\sigma_{z,B}-\sigma_{z,A}) (b^\dag-b)]
\end{equation}
with $\lambda_\pm=g/(\sqrt{2}\omega_\pm)$. The correspondingly diagonalised Hamiltonian reads
\begin{equation}\label{ising}
H'''=\frac{\omega'_0}{2}\left(\sigma_{z,A}+\sigma_{z,B}\right)+\frac{\chi}{2}\sigma_{z,A}\sigma_{z,B}+\omega_+a^\dag a+ \omega_-b^\dag b
\end{equation}
with $\chi\equiv {4g^2\lambda}/(\omega_-\omega_+)$.
The qubit part of Eq.~(\ref{ising}) has the form of an Ising Hamiltonian, demonstrating that the qubits come into interaction with each other through the coupled localized modes. As expected, this coupling constant goes to zero if the modes are decoupled ($\lambda=0$) or one of the qubits is detached from its local mode ($g=0$). Finally, as $T''T'T$ commutes with the free energy of the qubits, the dephasing part of Eq.~(\ref{dep}) is not affected by the transformations. Therefore, the dynamics in this transformed space is that of an Ising system subjected to dephasing with no influence from the modes. However, in the process of transforming the observables, the modes and qubits get in fact correlated. The system dynamics in the transformed space is governed by
\begin{equation}\label{dynamics}
\begin{aligned}
\frac{\partial\rho'''}{\partial t}=&-i[\omega_+a^\dag a+ \omega_-b^\dag b,\rho''']\\
&-\frac{i}{2}[\omega'_0\left(\sigma_{z,A}+\sigma_{z,B}\right)+\chi\sigma_{z,A}\sigma_{z,B},\rho''']\\
&+\frac{\gamma}{2}\left\{\left(\sigma_{z,A}\rho'''\sigma_{z,A}-\rho'''\right)+\left(\sigma_{z,B}\rho'''\sigma_{z,B}-\rho'''\right)\right\}.
\end{aligned}
\end{equation}
As this equation is diagonal in the common basis of the observables $\{\sigma_{z,A},\sigma_{z,B},a^\dag a,b^\dag b\}$, it is straightforward to solve it for any initial condition.

Here, we will explore the solution of this problem for a particular set of initial conditions which are useful to harness the non-Markovian character of the evolution of the qubits, and how this depends on the bosonic environment considered here. We will see that, as far as the dynamics of one of the qubits is concerned, 
the structure of the problem at hand is very rich. It includes bosonic modes, their cross coupling, the presence of a second qubit and Markovian dephasing. All such coherent and incoherent couplings compete to give rise to the features discussed in next sections. 
\section{Non-Markovianity under fully coherent conditions}
\begin{figure}[ht]\vspace{0.3cm}
\centering\includegraphics[width=0.8\columnwidth]{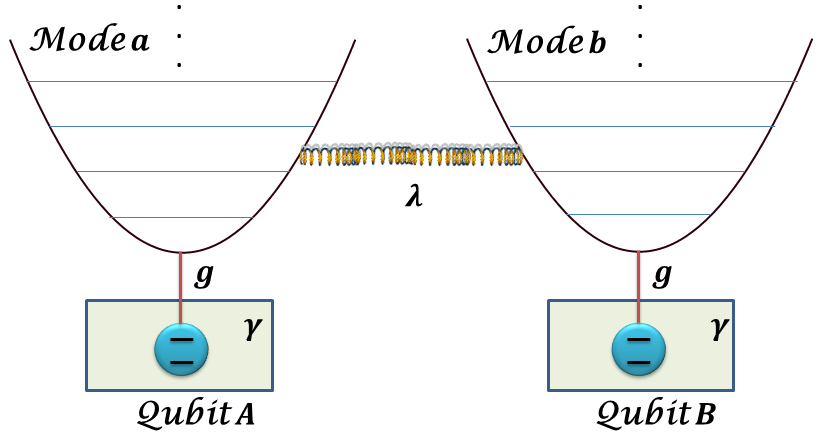}
\caption{(Color
  online) We depict the basic idea behind the investigation of non-Markovianity in this work. Qubit $A$ is subjected to Markovian dephasing $\gamma$ and is coupled to other subsystems directly through $g$ and indirectly thorough $\lambda$. Provided $g=0$, the time evolution of qubit $A$ will certainly be Markovian due solely to the $\gamma$ environment. Once $g$ is turned on, non-Markovian features may appear. The situation becomes even more interesting by introducing the inter-mode coupling.}
\label{ideia}
\end{figure}

From now on, we will focus on the non Markovianity of the evolution of qubit $A$. In particular, we would like to study the competition between the Markovian environment characterized by the $\gamma$-term in (\ref{dynamics}) and the presence of other quantum subsystems which influence qubit $A$. The general idea is depicted and explained in Fig.~\ref{ideia} and its caption.

According to \cite{nmb}, non Markovianity in open system dynamics of a qubit can be detected or inferred by considering the trace distance $D[\rho_1(t),\rho_2(t)]={\rm{Tr}}\,|\rho_1(t)-\rho_2(t)|/2$ between two evolved states, where $|A|=\sqrt{A^\dag A}$. The evolution process will be non-Markovian if there exists a pair of initial states $\rho_{1,2}(0)$ such that, after a time $t$, $\rho_{1,2}(t)$ will lead to $\sigma(t)>0$ where
\begin{eqnarray}
\sigma(t)=\frac{d}{dt}D[\rho_1(t),\rho_2(t)].
\end{eqnarray} 

In our case, qubit $A$ interacts with its environment (qubit $B$ plus dephasing bath) only by means of $\sigma_{z,A}$, as it can be seen from (\ref{dep}). For such cases, any pair of antipodal initial states living in the equatorial line in the Bloch sphere are expected to maximize $\sigma(t)$ \cite{max}. We checked this numerically. We will then consider $\rho_A(0)=|\pm\rangle_A\langle\pm|$, where $|\pm\rangle_A$ are eigenstates of $\sigma_{x,A}$ with eigenvalues $\pm 1$, respectively.  For qubit $B$ we will take a simple preparation which consists of letting it start from its ground state $\rho_B(0)=|g\rangle_B\langle g|$. For the modes, we will consider them in coherent states $|\alpha e^{i \theta}\rangle_a$ and $|\beta e^{i \varphi}\rangle_b$, with $\alpha, \beta, \theta$, and $\varphi$ real numbers. These states have already been generated in circuit-QED by driving the resonator with a microwave pulse with gaussian-shape \cite{coh}. In next section, we will use the solution for the pure coherent states 
obtained here to investigate the case of incoherent superpositions in a circle.

Let us then consider the qubits and modes to be initially prepared in the state
\begin{eqnarray}\label{in}
\rho_{\pm}(0)=|\pm\rangle_A\langle\pm|\otimes|g\rangle_B\langle g|\otimes|\alpha e^{i \theta}\rangle_a \langle\alpha  e^{i \theta}|\otimes|\beta e^{i \varphi}\rangle_b\langle\beta e^{i \varphi}|.
\end{eqnarray}
In order to solve Eq.~(\ref{dynamics}), we must first transform this state using the set $T''T'T$. The result is $\rho_{\pm}'''(0)= |\phi\rangle \langle \phi|$, where
\begin{eqnarray}
|\phi\rangle = \frac{e^{-2\lambda_-i{\rm{Im}}w_1^*}|e\rangle_A |g\rangle_B |z_1\rangle_a |w_2\rangle_b \pm e^{-2\lambda_+i{\rm{Im}}z_1^*}|g\rangle_A |g\rangle_B |z_2\rangle_a |w_1\rangle_b}{\sqrt{2}},\nonumber\\
\end{eqnarray}
with coherent states characterized by the complex numbers $z_1 = (\alpha e^{i \theta} + \beta e^{i \varphi} + 2 \xi)/\sqrt{2}$, $z_2 = (\alpha e^{i \theta} + \beta e^{i \varphi} + 2 \xi)/\sqrt{2} - 2 \lambda_+	$, $w_1 = (\beta e^{i \varphi} - \alpha e^{i \theta})/\sqrt{2}$ and $w_2 = (\beta e^{i \varphi} - \alpha e^{i \theta})/\sqrt{2} - 2 \lambda_-$. The corresponding evolved state, obtained using Eq.~(\ref{dynamics}), is thus
\begin{widetext}
\begin{eqnarray}
\rho_{\pm}'''(t)&=& \frac{1}{2} |e\rangle_A\langle e|\otimes|g\rangle_B\langle g|\otimes|z_1 e^{-i \omega_+ t}\rangle_a\langle z_1 e^{-i \omega_+ t}|\otimes|w_2 e^{-i \omega_- t}\rangle_b\langle w_2 e^{-i \omega_- t}|\nonumber\\
&& +\frac{1}{2} |g\rangle_A\langle g|\otimes|g\rangle_B\langle g|\otimes|z_2 e^{-i \omega_+ t}\rangle_a\langle z_2 e^{-i \omega_+ t}|\otimes|w_1 e^{-i \omega_- t}\rangle_b\langle w_1 e^{-i \omega_- t}|  \nonumber\\
&&\pm \frac{1}{2}e^{- \gamma t - i (\omega'_0 - \chi)t - i \sqrt{2} \lambda_- \left(\alpha \sin(\theta)-\beta \sin(\varphi)\right)- i \sqrt{2} \lambda_+ \left(\alpha \sin(\theta) + \beta \sin(\varphi)\right) }|e\rangle_A\langle g|\otimes|g\rangle_B\langle g|\otimes|z_1 e^{-i \omega_+ t}\rangle_a\langle z_2 e^{-i \omega_+ t}|\otimes |w_2 e^{-i \omega_- t}\rangle_b\langle w_1 e^{-i \omega_- t}| \nonumber \\
&&\pm \frac{1}{2}e^{- \gamma t + i (\omega'_0 - \chi)t +i \sqrt{2} \lambda_- \left(\alpha \sin(\theta)-\beta \sin(\varphi)\right)+i \sqrt{2} \lambda_+ \left(\alpha \sin(\theta) + \beta \sin(\varphi)\right) }|g\rangle_A\langle e|\otimes|g\rangle_B\langle g|\otimes|z_2 e^{-i \omega_+ t}\rangle_a\langle z_1 e^{-i \omega_+ t}|\otimes |w_1 e^{-i \omega_- t}\rangle_b\langle w_2 e^{-i \omega_- t}|.\nonumber\\	
\end{eqnarray}
\end{widetext}
In order to evaluate the non Markovianity of the evolution of qubit $A$, we must find its evolved state $\rho_{\pm}^A(t)$. We then need to transform back $\rho_{\pm}'''(t)$ to $\rho_{\pm}(t)$  and trace out qubit $B$ and the modes. By doing this, one obtains
\begin{equation}
\label{state}
\rho_{\pm}^A(t)=\frac{1}{2} \openone_A \pm (h(t) |e\rangle_A\langle g|\pm h.c.),
\end{equation}
where 
$h(t)=e^{\psi(t)}\langle \alpha_2|\alpha_1\rangle \langle \beta_2|\beta_1\rangle/2$
with 
\begin{eqnarray}
 \alpha_1 &=& (z_1 e^{- i \omega_+ t} - w_2 e^{- i \omega_- t} - 2 \lambda_-)/\sqrt{2} - \xi, \nonumber\\
 \alpha_2 &=& (z_2 e^{- i \omega_+ t} - w_1 e^{- i \omega_- t} + 2 \lambda_+)/\sqrt{2} - \xi, \nonumber\\
 \beta_1 &=& (z_1 e^{- i \omega_+ t} + w_2 e^{- i \omega_- t} + 2 \lambda_-)/\sqrt{2} - \xi,\nonumber\\
 \beta_2 &=& (z_2 e^{- i \omega_+ t} + w_1 e^{- i \omega_- t} + 2 \lambda_+)/\sqrt{2} - \xi,
\end{eqnarray}
and $\psi(t)=- \gamma t+i[(\chi-\omega'_0)t+\,\Gamma(t)]$ where
\begin{equation}
\begin{aligned}
\Gamma(t)&=4 [\lambda_+^ 2 \sin(\omega_+ t) - \lambda_-^ 2 \sin(\omega_- t) - \sqrt{2} \xi \lambda_+ \sin(\omega_+ t)]\nonumber \\ 
&+\sqrt{2}\alpha\sin(\theta)\{\lambda_+[\cos(\omega_+t)-1]+\lambda_-[\cos(\omega_-t)-1]\}\nonumber\\
&-\sqrt{2}\alpha\cos(\theta)[\lambda_+\sin(\omega_+t)+\lambda_-\sin(\omega_-t)]\nonumber\\
&+\sqrt{2}\beta\sin(\varphi)\{\lambda_+[\cos(\omega_+t)-1]-\lambda_-[\cos(\omega_-t)-1]\}\\
&-\sqrt{2}\beta\cos(\varphi)[\lambda_+\sin(\omega_+t)-\lambda_-\sin(\omega_-t)].
\end{aligned}
\end{equation}
We see that the coupling to the resonators and qubit $B$ directly affects the coherence of qubit $A$ through the decoherence factor $h(t)$. As a result, we expect the degree of non-Markovianity to be a function of $h(t)$. In fact, after evaluating the trace distance, one obtains $D[\rho_+^A(t),\rho_-^A(t)]=2|h(t)|$, which results in
\begin{eqnarray}\label{st}
\sigma(t)=e^{k(t)-\gamma t}f(t)
\end{eqnarray}
with
\begin{equation}
\label{gt}
k(t)=-4g^2\frac{(\lambda^2+\omega^2)(1-\cos[\lambda t]\cos[\omega t])-2\lambda\omega\sin[\lambda t]\sin[\omega t]}{(\lambda^2-\omega^2)^2},
\end{equation}
\begin{equation}\label{ft}
f(t)=\frac{\gamma (\omega^2-\lambda^2)-4g^2(\lambda\sin[\lambda t]\cos[\omega t]+\omega\cos[\lambda t]\sin[\omega t])}{\lambda^2-\omega^2}.
\end{equation}

The first interesting feature of $\sigma(t)$ given by Eq.~(\ref{st}) is its independence on $\alpha$ and $\beta$. 
This is a direct consequence of the fact that those amplitudes can be completely removed from the evolved state through a unitary time independent displacement. As a consequence, non-Markovianity of the qubit system can not depend on the information about the initial position in phase space of the coherent states for the modes. In the next Section, we will change the initial state of the modes by incoherently superimposing coherent states in a circle in phase space, and this will induce a dependence on the amplitude of the superposed states. The reason is that now information about amplitude of the coherent states can no longer be removed from the dynamics by means of a unitary transformation in the modes.

All our simulations are run using typical circuit-QED values~\cite{Schoelkopf,cqed,guo,expd} with $\omega_0/2\pi~\simeq5-10$ GHZ for CPB charge qubits and resonator frequencies $\omega/2\pi=10$ GHz~\cite{Schoelkopf}. For the qubit-resonator coupling constant, values of $g/2\pi\simeq10-100$ MHz are realistic, while dephasing rates for charge qubits as low as $\gamma/2\pi=0.3$ MHz have been measured \cite{expd}. As mentioned before, damping has been neglected in this first approach since the associated rates are much smaller than dephasing. The typical damping rate for the resonator is of a few KHz, and for charge qubits it can be made one order of magnitude smaller than $\gamma$~\cite{expd}. As for the inter-mode coupling strength, $\lambda/2\pi\simeq17$ MHz can be achieved with current experimental technologies \cite{guo}. Consequently, the conditions of strongly coupled modes, i.e. $g\approx\lambda$, can in principle be met.

\begin{figure}[t]
\centering\includegraphics[width=0.8\columnwidth]{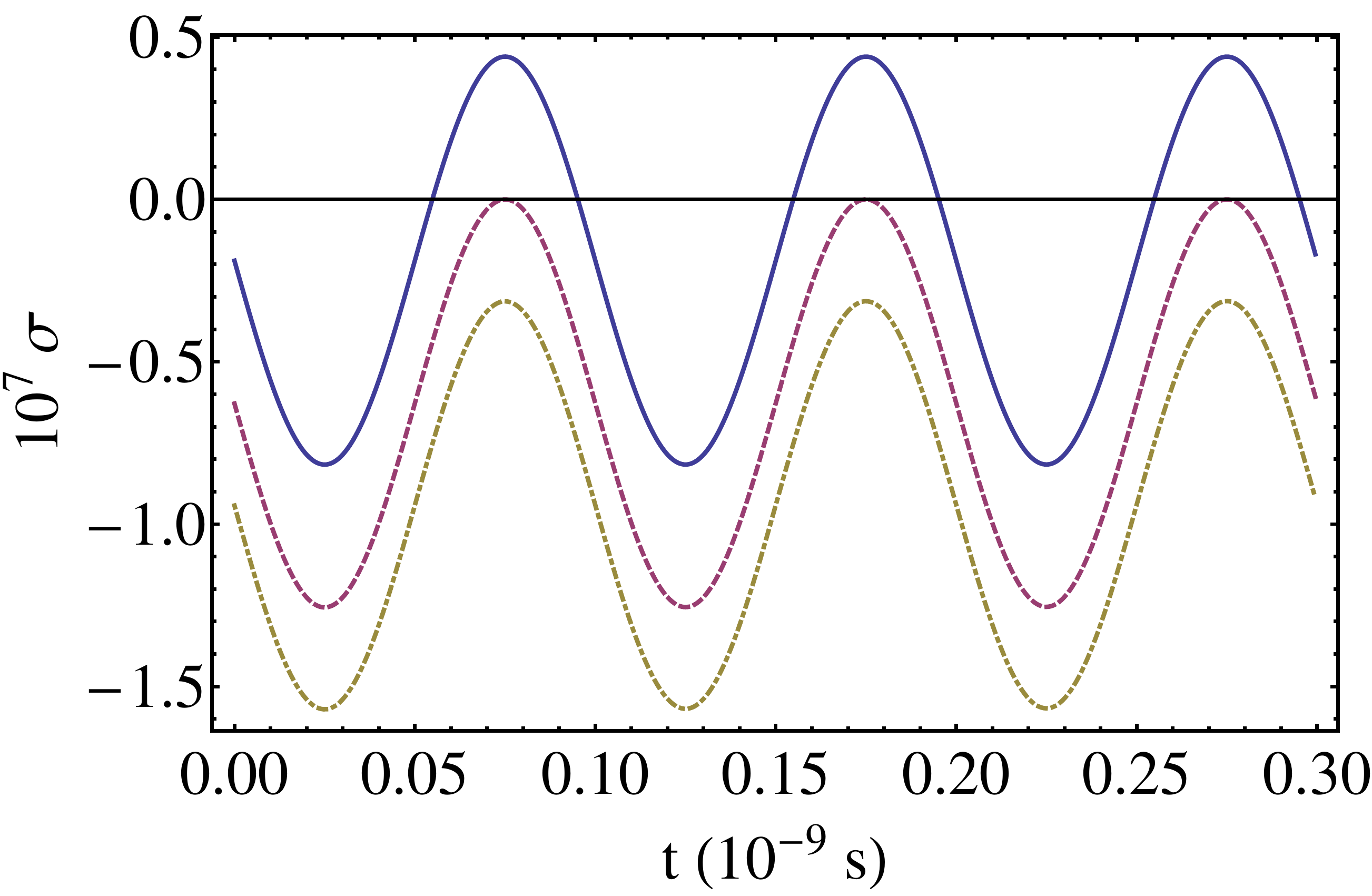}
\caption{(Color
  online) Dynamics of $\sigma(t)$ for various values of the dephasing rates and $\lambda=0$. We have taken $\gamma/2\pi=0.3$ MHz (solid line), $1.0$ MHz (dashed line), and $1.5$ MHz (dot-dashed one). As for the other parameters, we used $g/2\pi=50.0$ MHz and $\omega/2\pi=10.0$ GHz.}
\label{lambdazero}
\end{figure}

In Fig.~\ref{lambdazero}, we consider the role of dephasing $\gamma$ for the case $\lambda=0$ which corresponds to decoupled modes. It is very clear that by increasing the participation of the Markovian channel, it comes to a point where the non-Markovianity is unable to manifest. One could also fix the dephasing rate, and vary the spin-boson coupling $g$. The result will be exactly the opposite, i.e, above a threshold value for $g$, the dynamics becomes non-Markovian. From this, we can clearly see that there is a competition between the Markovian character of the dephasing environment, given by rate $\gamma$, and the highly non-Markovian local bosonic environment represented by the coupling constant $g$. A similar phenomenon of has been observed experimentally for the Ising model~ \cite{cbpf}.

\begin{figure}[t]
\centering\includegraphics[width=0.8\columnwidth]{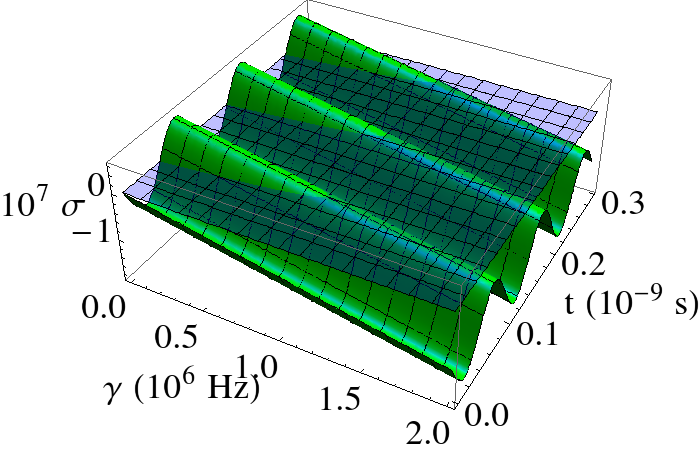}
\caption{(Color
  online) Dynamics of $\sigma(t)$ as a function of time and dephasing. We used $g/2\pi=50.0$ MHz and $\omega/2\pi=10.0$ GHz.}
\label{3D}
\end{figure}

From Eq.~(\ref{st}), it is easy to understand the nature of this threshold, which is physically due to the competition between the ``channels'' affecting the system at rates $\gamma$ and $g$. In order to see this, let us explicitly write $\sigma(t)$ for $\lambda=0$. According to Eq.~(\ref{st}), this reads
\begin{eqnarray}\label{sts}
\sigma(t)=-2[\gamma+4g^2\sin(\omega t)/\omega]e^{-2\gamma t-8g^2[1-\cos(\omega t)]/\omega^2}.
\end{eqnarray} 
In order for the qubit to follow a Markovian evolution, $\sigma(t)$ must be negative or null at all times. From Eq.~(\ref{sts}), we see that this will be the case provided that
\begin{eqnarray}\label{parabola}
\gamma>4g^2/\omega.
\end{eqnarray} 
This defines a parabolic boundary separating the Markovian and non-Markovian regimes in the parameter space formed by $\gamma$ and $g$. For the parameters considered in Fig.~\ref{lambdazero}, we can use Eq.~(\ref{parabola}) to obtain $\gamma/2 \pi>1.0$ MHz, which is precisely what is observed. In Fig.~\ref{3D}, one can clearly see the existence of a limit value of $\gamma$ above which $\sigma$ ceases to be positive for all times. This limit is given by (\ref{parabola}).

\begin{figure}[b]
\centering\includegraphics[width=0.8\columnwidth]{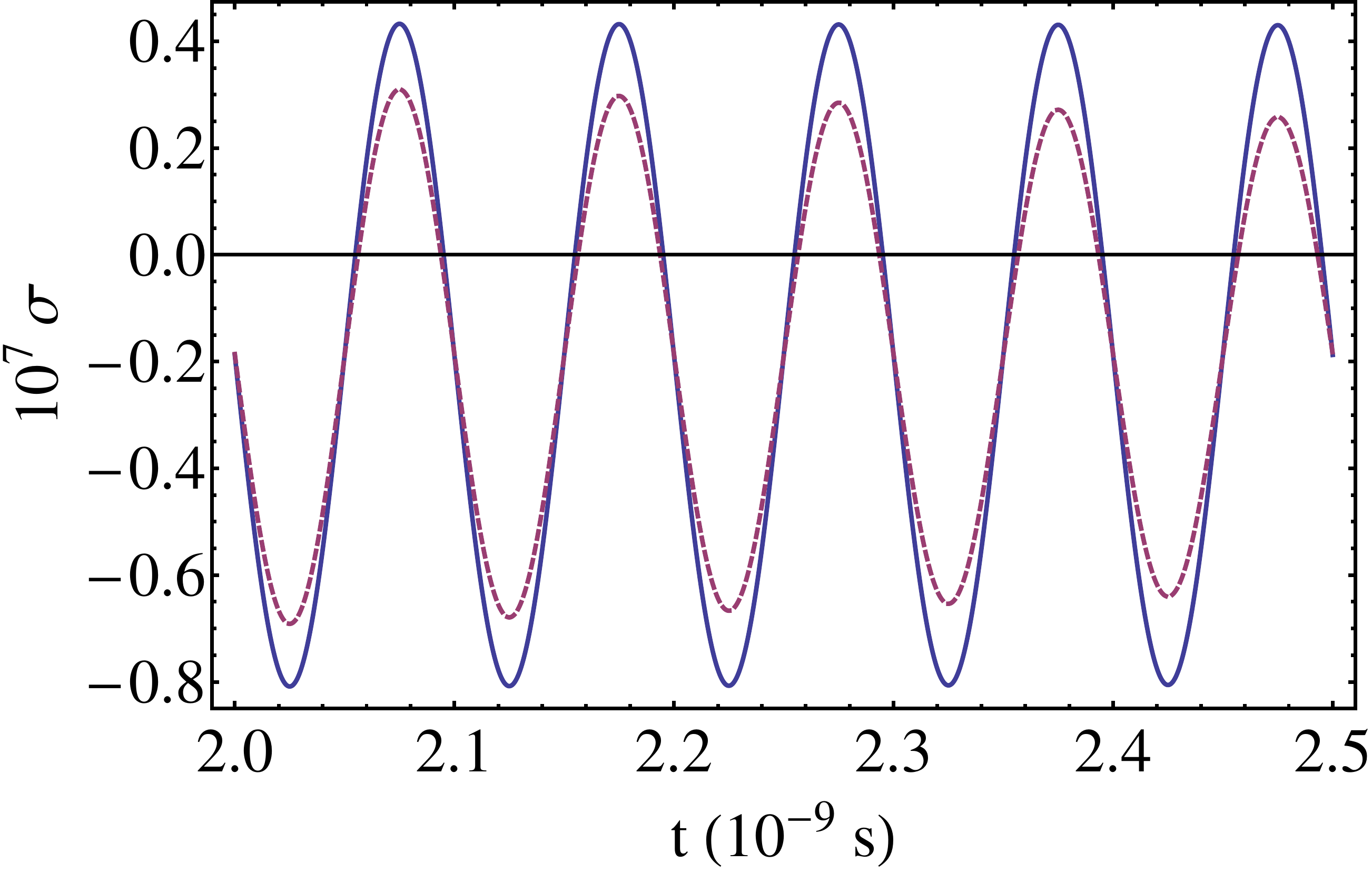}
\caption{(Color
  online) Dynamics of $\sigma(t)$ for various mode-mode coupling constants $\lambda$. We considered $\lambda/2\pi=10$ MHz (solid) and $\lambda/2\pi=50$ MHz (dashed). For the other parameters we used $\gamma/2\pi=0.3$ MHz, $g/2\pi=50.0$ MHz, and $\omega/2\pi=10.0$ GHz.}
\label{lambdanoventa}
\end{figure}

We consider now the effect of cross-coupling between the modes ($\lambda\neq0$). First, the effect of increasing $\gamma$ is still the progressive inhibition of backflow of information. In Figure \ref{lambdanoventa} we fix $\gamma$ and increase the coupling strength $\lambda$ betweem the modes. We can see that by increasing $\lambda$, non-Markovianity is also progressively diminished. This decreasing of the degree of non-Markovianity can be physically understood from fact that the mode coupled to qubit $A$ now becomes correlated with other quantum systems. This reduces its capability to get correlated, quantum mechanically, with qubit $A$, which in turn depletes the possibility to provide the backflow of information.
\section{Non-Markovianity under phase diffused bosonic modes}
Previously, we considered the modes to be prepared in pure coherent states and found that the trace distance and its time derivative became independent on the amplitudes $(\alpha,\beta)$ and phases $(\theta,\varphi)$ of the coherent states considered in the initial state (\ref{in}). Given such phase-independence, one could then think of using a mixture of iso-energetic coherent states with no phase coherence. Such mixture is constructed as
\begin{eqnarray}\label{dif}
\rho_a=\int_0^{2\pi}\frac{d\theta}{2\pi}\,|\alpha e^{i\theta}\rangle_a\langle\alpha e^{i\theta}|=e^{-|\alpha|^2}\sum_n\frac{|\alpha|^{2n}}{n!}|n\rangle_a\langle n|.
\end{eqnarray}
This state is central in the discussions about the quantum description of laser light and its ability to perform quantum information tasks~\cite{laser}. Both states, $|\alpha e^{i\theta}\rangle_a\langle\alpha e^{i\theta}|$ and $\rho_a$, have the same diagonal elements in the energy eigenbasis. However, the trace distance is not a linear function on the input states. Consequently, the use of mixtures of coherent states having the same energy might actually lead to different results. In fact, as we are going to see, the use of such mixed state brings about a dependence on the amplitudes $\alpha$ and $\beta$, which marks a substantial difference with respect to the pure state case. We now consider
\begin{eqnarray}\label{nis}
\rho_{\pm}(0)=|\pm\rangle_A\langle\pm|\otimes|g\rangle_B\langle g|\otimes\rho_a\otimes\rho_b,
\end{eqnarray} 
with $\rho_b$ given by (\ref{dif}) upon changing $\alpha$ to $\beta$ and $\theta$ to $\varphi$. We can use the results of the previous section to evolve the states, and the time derivative of the trace distance is found to be
\begin{eqnarray}\label{sm}
\sigma_{mix}(t)&=&e^{k(t)-\gamma t}\{-\sqrt{2}g^2\alpha\, J_0[\beta F_1(t)]\,J_1[\alpha F_2(t)]\,G_1(t)\nonumber\\ &&-\sqrt{2}g^2\beta\, J_0[\alpha F_2(t)]\,J_1[\beta F_1(t)]\,G_2(t)\nonumber\\ &&+J_0[\beta F_1(t)]\,J_1[\alpha F_2(t)]\,G_3(t)\},
\end{eqnarray}
where $J_n(x)$ are Bessel functions of order $n$, $k(t)$ is given in Eq.~(\ref{gt}), and
\begin{widetext}
\begin{eqnarray}
F_1(t)&=&2\sqrt{2}g\sqrt{\frac{\,3\lambda^2+\omega^2+(\lambda^2-\omega^2)\cos[2\lambda t]-4\lambda\,(\lambda\cos[\lambda t]\cos[\omega t]+\omega\sin[\lambda t]\sin[\omega t])}{(\lambda^2-\omega^2)}},\nonumber\\
F_2(t)&=&2\sqrt{2}g\sqrt{\frac{\,\lambda^2+3\omega^2+(\omega^2-\lambda^2)\cos[2\lambda t]-4\omega\,(\omega\cos[\lambda t]\cos[\omega t]+\lambda\sin[\lambda t]\sin[\omega t])}{(\lambda^2-\omega^2)}},\nonumber\\
G_1(t)&=&\frac{8 \sqrt{2}\cos[\lambda t](\lambda\sin[\lambda t]-\omega\sin[\omega t])}{(\lambda^2-\omega^2)F_2(t)},\nonumber\\
G_2(t)&=&\frac{8 \sqrt{2}\sin[\lambda t](-\cos[\lambda t]+\cos[\omega t])}{(\lambda^2-\omega^2)F_1(t)},\nonumber\\
G_3(t)&=&-\gamma-\frac{2g^2(\omega-\lambda)\sin[(\omega-\lambda)t]}{(\omega-\lambda)^2}-\frac{2g^2\sin[(\omega+\lambda)t]}{\omega+\lambda}.
\end{eqnarray} 
\end{widetext}

\begin{figure}[b]
\centering\includegraphics[width=0.8\columnwidth]{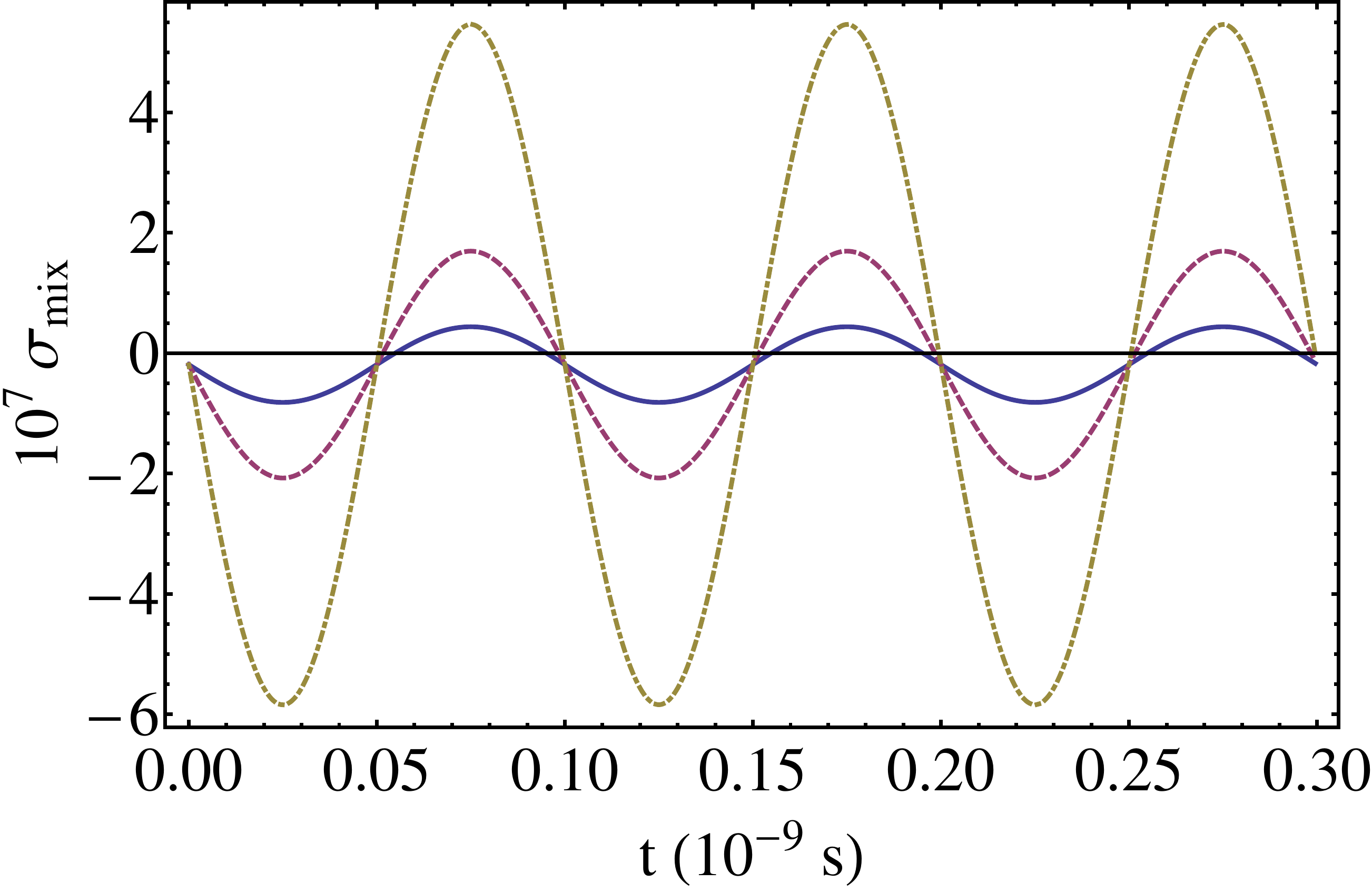}
\caption{(Color
  online) Dynamics of $\sigma_{mix}(t)$ obtained considering the initial state in Eq.~(\ref{nis}) for various amplitudes $\alpha$ and $\beta$. We took $\alpha=\beta=0$ (solid), $\alpha=\beta=1$ (dashed), and $\alpha=\beta=2$ (dot dashed). For the other parameters we used $\gamma/2\pi=0.3$ MHz, $g/2\pi=50.0$ MHz, and $\omega/2\pi=10.0$ GHz}
\label{varamp}
\end{figure}

Let us now focus our attention on the case $\alpha=\beta$. If these amplitudes are null, it is not difficult to see that (\ref{sm}) reduces to (\ref{st}) as expected for arbitrary $\lambda$. As we did before, let us start  the analyzes by considering the case of decoupled modes ($\lambda=0$). The effect of increasing the amplitudes of the coherent states are presented in Fig.~\ref{varamp}. It is noticeable that the effect of increasing the amplitudes (energy) of the modes, the non-Markovianity increases. As the entropy (mixedness) of the initial state increases with $\alpha$ and $\beta$, one could, as a first guess, expect that the non-Markovianity arising from the coupling to the modes would decrease as the amplitudes increase. However, our results show that for this particular mixtures of coherent states in a circle, the opposite happens. The total elimination of the off diagonal elements due to the integral over equally weighted phases, not only made the results dependent on the amplitudes, but also 
brought about this particular effect.

\begin{figure}[t]
\centering\includegraphics[width=0.8\columnwidth]{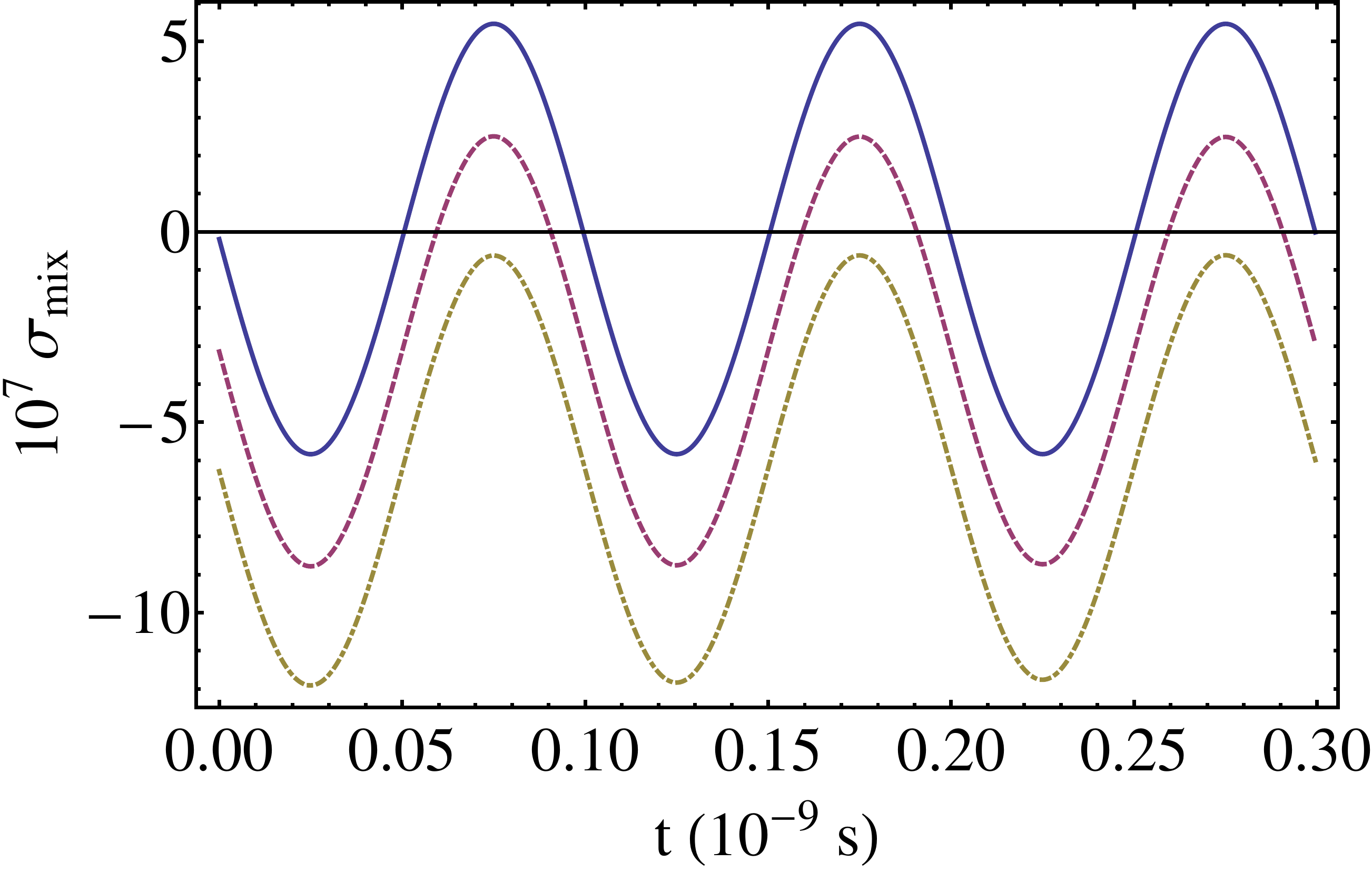}
\caption{(Color
  online) Dynamics of $\sigma_{mix}(t)$ obtained considering the initial state in Eq.~(\ref{nis}) for various dephasing rates and decoupled modes $\lambda=0$. We considered $\gamma/2\pi=0.3$ MHz (solid), $\gamma/2\pi=5.0$ MHz (dashed), $\gamma/2\pi=10.0$ MHz (dot dashed). For the other parameters we used $g/2\pi=50.0$ MHz and $\omega/2\pi=10.0$ GHz.}
\label{ampgam}
\end{figure}

For the case of coupled modes ($\lambda\neq0$), the behavior for fixed $\alpha$ and $\beta$ is similar to the one found in previous section. By increasing $\lambda$, non-Markovianity tends to decrease. Finally, for this initial mixture of coherent states, there is again a competition between $g$ and $\gamma$. The results are shown in Fig.~\ref{ampgam}. By increasing $\gamma$, it comes to a point where the dynamics is fully Markovian. However, given the complicated dependence of $\sigma_{mix}(t)$ on $g$ and $\gamma$, it is not possible now to obtain an analytical formula for the Markovianity boundary.

As a final remark, we would like to talk about the experimental challenges involved in the characterization of non-Markovianity. The full experimental evaluation of the non-Markovianity measures requires the tomography of states or processes. This follows from the fact that non-Markovianity is not, in general, pinpointed by an observable (like level populations). These facts forbid, in most cases, that a reliable signature of non-Markovianity can be inferred from directly accessible quantities in an experiment. For a specific model, it has been shown that non-Markovianity is accompanied by violation of macrorealism \cite{cbpf}. This kind of work linking non-Markovianity to fundamental issues of quantum mechanics \cite{us} is a very exciting field with great importance for the understanding of the classical/quantum boundary. 

\section{Conclusion} 
We have assessed the problem of non-Markovianity characterization in a specific circuit-QED setup consisting of two qubits, each of them locally coupled to a bosonic mode. The modes can be controllably coupled to each other through common interaction with a third qubit. We have solved the corresponding model exactly and studied non-Markovianity for the qubits from the point of view of information backflow from environment to qubit. For modes prepared in pure coherent states or mixtures of equally weighted coherent states with fixed energy, we found analytical expressions for the quantifier of information backflow, which is the trace distance. The general effect of having Markovian dephasing acting on the qubits is the existence of a threshold of Markovinity i.e., a lower bound for the dephasing rate, above which the evolution is purely Markovian. For decoupled modes, we found the analytical dependence of this lower bound on parameters of the system.

Although the degree of non-Markovianity for initial pure coherent states are independent on the amplitude and phase of these states, for a mixture of coherent states in a circle, the result becomes actually dependent on the amplitudes. Surprisingly, the bigger the amplitudes, the more non-Markovian the qubit dynamics becomes. Our work contributes to the study and control of open quantum systems by presenting, in a versatile setup, the complete diagonalization of the open system dynamics and an comprehensive characterization of non-Markovianity.   
\begin{acknowledgments}
P.C.C. wishes to thank FAPESP for the support through Grant. No. 2012/12702-7. M.P. thanks the John Templeton Foundation (grant 43467) and the UK EPSRC for a Career Acceleration Fellowship and a grant awarded under the ``New Directions for Research Leaders" initiative (EP/G004579/1). FLS and MP are supported by the CNPq ``Ci\^{e}ncia sem Fronteiras'' programme through the ``Pesquisador Visitante Especial'' initiative (grant nr. 401265/2012-9). F.L.S. acknowledge participation as members of the Brazilian National Institute 
of Science and Technology of Quantum Information (INCT/IQ). F.L.S. also acknowledges partial support from CNPq under grant $308948/2011-4$. 
\end{acknowledgments}


\end{document}